# A T cell equation as a conceptual model of T cell responses for maximizing the efficacy of cancer immunotherapy

Haidong Dong[1,2*], Yiyi Yan[3,4], Roxana S. Dronca[3], and Svetomir N. Markovic[1,4]

[1]*Department of Immunology, Mayo Clinic, Rochester, MN, USA*
[2]*Department of Urology, Mayo Clinic, Rochester, MN, USA*
[3]*Division of Medical Oncology, Mayo Clinic, Rochester, MN, USA*
[4]*Division of Hematology, Mayo Clinic, Rochester, MN, USA*

## Abstract

Following antigen stimulation, the net outcomes of a T cell response are shaped by integrated signals from both positive co-stimulatory and negative regulatory molecules. Recently, the blockade of negative regulatory molecules (i.e. immune checkpoint signals) demonstrates promising therapeutic effects in treatment of human cancers, but only in a fraction of cancer patients. Since this therapy is aimed to enhance T cell responses to cancers, here we devised a conceptual model by integrating both positive and negative signals in addition to antigen stimulation that can evaluate strategies to enhance T cell responses. A digital range of adjustment of each signal is formulated in our model for prediction of a final T cell response. Our model provides a rational combination strategy for maximizing the therapeutic effects of cancer immunotherapy.

## Introduction

T cell response to antigen stimulation is a tightly controlled process. Recent clinical trials have demonstrated that unleashing T cell responses to cancers could be an effective approach in the treatment of human malignancies [1-3]. The identification of immune checkpoint molecules in tumor immune evasion greatly contributes to the development of therapeutics aimed to block immunosuppressive mechanisms in order to enhance antitumor T cell immunity [4,5]. Although the clinical outcomes of immune checkpoint blockade are promising, the low efficiency and potential adverse effects remain as major challenges. Combination therapy among different immune checkpoint targets or other therapies (i.e. chemotherapy, targeting therapy or radiotherapy, etc.) is speculated to increase the efficacy of cancer immunotherapy [6,7]; however, the rationale of optimized combination is still lacking. Given the complex responses of T cells that are regulated by a battery of signals at different stages of activation and differentiation [8], a conceptual model is needed to design an effective combined therapy that would maximize the therapeutic effects of each components of a regimen. Here, we devised a conceptual model of T cell responses that can be used to predict an outcome of a T cell response according to changes of positive or negative regulatory signals. Importantly, based on this equation, our model provides a rationale for synergistic treatment combination aimed to decrease resistance and maximize T cell responses against cancers.

## Two signal theory in shaping T cell responses

Antigen stimulation initiates a T cell response through the T Cell Receptor (TCR); however, the net outcome of a T cell response (activation, anergy, or tolerance) to this antigen is regulated by two additional signals, i.e. costimulation (CD28) or checkpoint (CTLA-4 or PD-1, etc.),which are integrated with the TCR signaling pathway [9]. To represent a T cell response that is initiated by antigen via TCR engagement and regulated by integrated positive or negative signals, we present an equation as below:

$$R = \frac{P*T}{N*T+1}$$

In this equation, R is for Response; T is for TCR signal; P is for Positive costimulation signals (like CD28 or signal 3 cytokines IL-2 or IL-12); N is for Negative checkpoint signals (like CTLA-4, PD-1 or IL-10). According to this equation, we defined that when R=1, a T cell response is turned on; when R=0, a T cell response is turned off; when R>1, a T cell response is enhanced; when R<1, a T cell response is deterred or in a tolerogenic status. In the following sections, we will give several examples of different outcomes of a T cell response based on the integration of TCR signals along with positive or negative regulatory signals in this mathematical model to see how our equation would predict a T cell response. The relationship of N or P with T is defined as N*T or P*T, rather than N+T or P+T according to the proximal signaling integration of N or P with TCR signals.

Besides checkpoint molecules that are directly integrated within TCR signaling pathway, other immune regulatory systems work in parallel with TCR signals to control T cell responses. These other regulatory mechanisms include, but are not limited to, regulatory T cells (Treg) or myeloid-derived suppressor cells (MDSCs) [10]. To include these parallel rather than proximal signals in regulation of TCR signals in addition to N, we used a constant number "1" as an inclusive and independent parameter in our T cell equation, and defined their relation with TCR using "+" to present the additional regulations.

**T cell response is dependent on antigen stimulation though the TCR**

First of all, this equation should be able to predict the fundamental role of antigen stimulation of TCR in the T cell response. Actually, in the absence of antigen stimulation and without TCR signals, i.e. T=0, then R will be 0, and there is no T cell response (**Calculation 1**).

$$Calculation\ 1: R = \frac{P*0}{N*0+1} = \frac{0}{0+1} = 0$$

This equation also explains why positive or negative signals *alone* do not have any effects on T cell response in the absence of TCR stimulation, since when T=0, R will always be 0 whether P or N is 1 or not.

**T cell response is dependent on costimulation and regulated by immune checkpoint signals**

It has been established that a full activation (response) of T cells is dependent on the presence of costimulation, i.e. CD28 engagement [11,12]. In the absence of costimulation (when P=0), R will always be 0, though there is a TCR stimulation (T=1) (**Calculation 2**). The outcome of calculation 2 explains T cell anergy [13], i.e. a mere TCR stimulation is not able to initiate a full T cell response.

$$Calculation\ 2: R = \frac{0*1}{0*1+1} = \frac{0}{0+1} = 0$$

When there is a costimulation signal (P=1), a full T cell response will be generated (R=1) (**Calculation 3**), as long as negative signals are absent (N=0).

$$Calculation\ 3: R = \frac{1*1}{0*1+1} = \frac{1}{0+1} = 1$$

If a negative signal is present (N=1), R will be 0.5, which is less than 1, indicating a deferred T cell response or T cell tolerance (**Calculation 4**).

$$Calculation\ 4: R = \frac{1*1}{1*1+1} = \frac{1}{1+1} = 0.5 < 1$$

According to Calculation 4, T cell tolerance is established by negative signals via immune checkpoint molecules. Calculation 4 also suggests that although both TCR stimulation and positive regulatory signals (costimulation) are present, there is no guarantee that a full T cell response can be generated due to immune regulatory mechanisms (N+1). Thus, our equation demonstrates a critical role of negative signals (immune checkpoints) in restraining the T cell response, which could be crucial in order to prevent pathology caused by any ongoing or unlimited T cell responses. We acknowledge that our T cell equation reflects how naïve T cells are primed to become activation, while the re-activation of memory T cells could be independent of co-stimulation (i.e. P signal).

**Breaking T cell tolerance and enhancing a T cell response**

As shown in Calculation 4, the presence of negative signals or immune checkpoints significantly compromises the generation of a full T cell response initiated by TCR stimulation in the presence of costimulation. In order to enhance T cell response or break a T cell tolerance, we have to increase the strength of either costimulation or TCR stimulation. To that end, if we increase costimulation P to 2, and keep the others at the same levels (T=1, N=1), we will have R = 1 (**Calculation 5**), suggesting a T cell response can be restored through the increase of costimulation. This calculation is in line with an early observation that introduction of CD28 costimulation (positive signals) enhances T cell response [11,12] or introduction of B7 (CD80/CD86) molecules into tumor cells results in a strong antitumor response in vivo [14].

$$Calculation\ 5: R = \frac{2*1}{1*1+1} = \frac{2}{1+1} = 1$$

However, a mere increase of TCR stimulation (let T=2) cannot restore a T cell response in the presence of negative signals (when N=1) (**Calculation 6**). This outcome may explain some preclinical and clinical observations showing strong antigenicity (e.g. high affinity antigen peptides) alone did not initiate a strong T cell response and failed to generate a protective T cell immunity. Occasionally, some high affinity antigen peptides may overcome the requirement of co-stimulation in activation of T cells, however, most of the cases involved activation of memory or memory-like T cells.

$$Calculation\ 6: R = \frac{1*2}{1*2+1} = \frac{2}{2+1} = 0.67 < 1$$

Next we examined to what degree a reduction of negative signals would be required to restore or enhance a T cell response. According to **Calculation 7**, when N is in a range of 0.1 to 0.9, R will always be less than 1, suggesting partial reduction of negative signals is not enough to restore a T cell response. As indicated from calculation 3, only a complete blockade or absence of negative signals, i.e. when N=0, a full T cell response can be achieved. This result underscores the strategy currently used in treatment of human cancer by a complete blockade of immune checkpoints (PD-1 or CTLA-4) in order to achieve objective clinical responses. Actually, the combination of anti-PD-1 and anti-CTLA-4 treatment achieved higher response rather than either alone [15].

$$Calculation\ 7: R = \frac{1*1}{0.1\,or\,0.9*1+1} = \frac{1}{0.1\,or\,0.9+1} = 0.9\,or\,0.5 < 1$$

However, since a complete blockade of negative signals only can be achieved in a fraction of cancer patients, and in most situations negative signals can only be partially reduced, additional approaches are needed to restore or increase a T cell response. To that end, if negative signals are partially reduced (let N=0.5), our equation suggests a partial increase of costimulation (when P=1.5) will be able to restore a T cell response (**Calculation 8**).

$$Calculation\ 8: R = \frac{1.5*1}{0.5*1+1} = \frac{1.5}{0.5+1} = 1$$

In order to enhance T cell response (i.e. to let R>1), a double increase of costimulation (let P=2) is needed as shown in **Calculation 9** if the negative signals are also partially reduced (N=0.5).

$$Calculation\ 9: R = \frac{2*1}{0.5*1+1} = \frac{2}{0.5+1} = 1.3 > 1$$

It could be very challenging, if not impossible, to have a double increase of costimulation in order to enhance a T cell responses, for example, to have poorly immunogenic tumor cells express B7 costimulatory ligand [14] or to provide additional costimulation signals directly to T cells (e.g. 41BB stimulation) [16]. Alternatively, a combination of partially increased TCR stimulation and costimulation (T=1.5; P=1.5) with partially decreased negative signals (N=0.5) will be able to give an enhanced T cell response(R=1.29 >1) (**Calculation 10**). This calculation indicates that a synergistic combination can be achieved by integrating a suboptimal increase of TCR and costimulation and suboptimal decrease of negative signals (e.g. partial immune checkpoint blockade) in order to enhance T cell responses.

$$Calculation\ 10: R = \frac{1.5*1.5}{0.5*1.5+1} = \frac{2.25}{0.75+1} = 1.29 > 1$$

## Discussion

Here we present a T cell equation as a conceptual model that can be used to predict the net outcomes of T cell responses by integrating both regulatory and stimulatory signals along with antigen stimulation. The T cell equation(R=P*T/[N*T+1]) gives a digital range (0.1-0.9 or 1-2) of adjustment for each regulatory or stimulatory signal T cells may receive during antigen stimulation. As predicted from Calculation 10, a synergistic combination is generated by integrating each signal when each signal can only be adjusted in a suboptimal condition due to practical limitations. The predication of our equation underscores the significance of current clinical efforts in seeking synergistic combination treatment of human cancers in order to decrease drug resistance and to increase the efficacy of cancer immunotherapy.

Our model predicts that simply increasing TCR stimulation is not enough to increase T cell responses due to the regulation of immune checkpoints at least in priming naïve T cells. In line with this predication, objective cancer responses have not been achieved in clinical trials with several tumor antigens that have strong antigenicity [17]. As predicted by our equation, if we combine tumor antigen peptides with immune adjuvants that are used to increase costimulation, i.e. to increase the expression of stimulatory molecules by antigen presenting cells, such tumor vaccine formulations are able to generate tumor antigen specific T cell responses [18]. However, since some adjuvants have the potential to increase the expression of immune checkpoint molecules [19], the therapeutic effects of tumor antigen vaccine are compromised due to immune regulatory mechanisms. To maximize the therapeutic effects of tumor vaccine, our calculations 9 and 10 suggest that components capable of increasing costimulation or decreasing negative regulatory signals, or both, should be integrated in an optimal formulation for tumor vaccines [20].

Our T cell equation clearly presented the significance of the immune checkpoint in controlling the T cell response. This prediction is echoed by recent successful treatments of some human cancers with the immune checkpoint blockade strategy (CTLA-4 or PD-1) that aimed to restore or enhance antitumor T cell immunity [21,22]. Interestingly, to achieve more efficient reduction of negative signals as shown in our equation to reduce N value as close as possible to 0, a combined therapy of both PD-1 and CTLA-4 therapy has been approved by FDA to gain a synergy effect in treatment of metastatic melanoma [15]. However, this combined blockade of immune checkpoints might increase the risk of enhanced adverse effects in some patients. As predicted by the T cell equation, adjustment of other stimulatory or regulatory signals should be considered in order to gain a safer and stronger antitumor T cell response. The field of cancer immunotherapy has moved from an era of empirical combinations to one of rational design by considering the compatibility of each regulatory or stimulatory mechanisms [7]. To that end, the predication of our conceptual model of T cell response provides a rationale to design a synergistic combination that takes into count each of the major factors that work together to affect the net outcome of a T cell response. As predicted by Calculation 10, a synergistic combination can be achieved by integrating sub optimally adjusted stimulatory or regulatory signals in order to enhance T cell responses in cancer patients.

Our T cell equation, as a conceptual model, is not designed for dose estimation or calculation in application of a particular regimen of cancer immunotherapy, but rather our model may predict a final outcome based on the signal strength a regimen may bring in. Since no defined dose-response has been established in cancer immunotherapy, i.e. the highest dose is not always the optimal dose, our model provides a way to evaluate how signals (positive or negative) are integrated for achieving a maximal effect in promoting T cell responses. Based on our T cell equation, a level of signal strength can be determined (for example to set N=0.5). Accordingly, the actual dose (concentration) of a regimen (antibody used to block immune checkpoint) will be determined by selecting a dose that would lead to 50 percent reduction of negative signals. The actual effects of 50 percent reduction of negative signals in cancer treatment eventually will be evaluated by objective biomarkers or clinical responses.

Some chemotherapy drugs cause immunogenic cell death (ICD) in tumor cells, such as doxorubicin, mitoxantrone, oxaliplatin, and cyclophosphamide. Accumulating clinical data indicate that activation of adaptive immune responses induced by immunogenic cell death is associated with improved disease outcome in cancer patients [23]. According to our T cell equation, the ICD of tumors likely contribute to the increase of T (TCR) signals by releasing more immunogenic tumor antigens, and to the increase of P (costimulation) signals by releasing a series of immunostimulatory damage-associated molecular patterns (DAMPs), so called natural adjuvants [24], that promotes antigen presentation and T cell priming [25]. In combination with immune checkpoint inhibitors (like anti-PD-1) that aim to reduce N (negative) signals, it is predictable that chemotherapy drugs that cause ICD can achieve additive or synergistic clinical activity by coincidently increasing P and T and decreasing N according to our T cell equation.

Taken together, our T cell equation provides a conceptual model of T cell responses for designing synergistic treatment combinations aimed to defuse resistance and maximize T cell responses against cancers. Our T cell equation indicates that a combined therapeutic formula should include approaches capable of increasing tumor antigen stimulation and costimulation, and at the same time, reducing or blocking immune checkpoint signals.

### Author contribution

H.D. conceived the concept, designed the key T cell equation, and wrote the main manuscript text. Y.Y., R.D., and S.M contributed to the discussion of clinical implications of this equation and revised versions of some equations. All authors reviewed the manuscript.

### Acknowledgements

Authors have no competing financial interests. Financial Support: Mayo Foundation (Dong), and in part by NCI R21 CA197878 (Dong/Dronca) and NIH/NIAID R01 AI095239 (Dong).